\documentclass{WileyMSP-template}

\begin{document}
\pagestyle{fancy}

\title{ Superconductive Sodalite-like Clathrate Hydrides MXH$_{12}$ with Critical Temperatures of near 300 K under Pressures}
\maketitle


\author{Yuxiang Fan}
\author{Bin Li*}
\author{Cong Zhu}
\author{Jie Cheng}
\author{Shengli Liu*}
\author{Zhixiang Shi}

\dedication{}

\begin{affiliations}
Bin Li\\
School of Science, Nanjing University of Posts and Telecommunications, Nanjing 210023, China\\
Jiangsu Provincial Engineering Research Center of Low-Dimensional Physics and New Energy, Nanjing 210023, China\\
Email Address: libin@njupt.edu.cn

Yuxiang Fan, Jie Cheng, Shengli Liu\\
School of Science, Nanjing University of Posts and Telecommunications, Nanjing 210023, China\\
Email Address: liusl@njupt.edu.cn

Cong Zhu\\
College of Electronic and Optical Engineering, Nanjing University of Posts and Telecommunications, Nanjing 210023, China\\

Zhixiang Shi\\
School of Physics, Southeast University, Nanjing 211189, China\\

\end{affiliations}

\begin{abstract}
We designed and investigated a series of ternary hydride compounds MXH$_{12}$ crystallizing in the cubic $Pm\overline{3}m$ structure as potential rare-earth and alkaline-earth superconductors. First-principles calculations were performed on these prospective superconductors across the pressure range of 50-200 GPa, revealing their electronic band structures, phonon dispersions, electron-phonon interactions, and superconducting properties. Several compounds were identified as dynamically stable, with ScYbH$_{12}$ and LuYbH$_{12}$ remaining stable at 70 GPa, and ScLuH$_{12}$ at 100 GPa. Notably, Eliashberg theory and electron-phonon coupling calculations predict CaLuH$_{12}$ to exhibit a remarkable $T_{c}$ of up to 294 K at 180 GPa. These findings unveil ternary hydrides as a promising class of high-temperature superconductors and provide insights for achieving superconductivity at lower or ambient pressures through material design and exploration.
\end{abstract}


\section{Introduction}

First-principles calculations have shown that hydrogen can transition into a metallic state at extremely high pressures exceeding 450 GPa, and in this metallic phase, it is predicted to exhibit superconductivity with a remarkably high superconducting transition temperature ($T_c$) of 242 K  \cite{1, 2}. Achieving the exceptionally high pressures required for metallization of pure hydrogen is considered an extreme experimental challenge \cite{3}. However, the incorporation of other elements with hydrogen can effectively 'chemically pre-compress' it \cite{4}, leading to the formation of compounds that exhibit high superconducting transition temperatures at significantly lower pressures compared to pure hydrogen. The pursuit of room-temperature superconductivity has drawn significant attention to hydrogen-rich compounds  \cite{5, 6, 7, 8, 9}, which have emerged as a promising research area. Many of these compounds are theoretically predicted to exhibit remarkably high superconducting transition temperatures well above 100 K, a projection that has been corroborated by subsequent experimental studies.  An especially noteworthy case is the experimental observation of superconductivity in H$_3$S with a $T_c$ of approximately 203 K at a pressure of 155 GPa \cite{10, 11, 12}.

Hydrides with a cage-like crystal structure have been identified as a promising class of materials for achieving high superconducting transition temperatures, leading to extensive research efforts focused on investigating hydrides exhibiting such structural features. Among the extensively studied systems, the rare-earth (RE) superhydrides, featuring cage-like structures such as REH$_6$, REH$_9$, and REH$_{10}$, are predicted to exhibit remarkably high superconducting transition temperatures. Notable examples include YH$_6$ \cite{13}, YH$_9$ \cite{14, 15}, YH$_1$$_0$ \cite{16}, LaH$_1$$_0$ \cite{17}, ScH$_6$ \cite{18}, LuH$_6$ \cite{19}, CeH$_9$ \cite{20}, CeH$_1$$_0$ \cite{21} and LaH$_1$$_0$ \cite{22}. A prominent structural feature found in many alkaline-earth and rare-earth metal superhydrides is the clathrate hexahydride $Im\overline{3}m$-XH$_6$ (X = Ca, Lu, Y). In this clathrate structure, hydrogen atoms occupy all the tetrahedral voids of a body-centered cubic ($bcc$) lattice, forming weakly covalent bonds within the cages. The metal atom X is situated at the center of the H$_{24}$ cage and acts as an electron donor, facilitating electron pairing and potentially contributing to the high superconducting transition temperatures observed in these materials \cite{19, 23, 24, 25, 26}. Density functional theory (DFT) calculations predict that the $T_{c}$ of CaH$_6$ is 235 K at 150 GPa \cite{26}, and experimental synthesis of CaH$_6$ with a sodalite-like H$_2$$_4$ cage has  confirmed a $T_c$  of 215 K at 172 GPa. Among the known XH$_6$ clathrate hydrides, the rare-earth hydride LuH$_6$ exhibits the highest $T_c$ of 273 K at 100 GPa, approaching the freezing point of water.

Incorporating additional elements into binary hydrides to form ternary hydrides is a key strategy for enhancing the superconducting transition temperature or lowering the pressure required to stabilize the superconducting phase. Li elements were incorporated into MgH$_1$$_6$ to fill the antibonding orbitals of H$_2$ molecular units, leading to the design of Li$_2$MgH$_1$$_6$, which exhibits the highest superconducting transition temperature ($T_c$ = 473 K) among ternary hydrides at a pressure of 250 GPa  \cite{27}. Researchers have discovered a range of hydrogen-rich materials exhibiting high-temperature superconductivity in a class of cage hydrides infused with rare-earth or alkaline-earth elements possessing $f$-shell electrons. Examples of such materials include $Fd\overline{3}m$-LaYH$_{10}$ with $T_c$ = 253 K at 183 GPa \cite{28}, $Fd\overline{3}m$-CaYH$_{12}$ with $T_c$ = 258 K at 200 GPa \cite{23}, and $P6_3/mmc$-(La, Ce)H$_{9-10}$ with $T_c$ = 176 K at 100 GPa \cite{29}. Metal elements with larger atomic radii induce chemical precompression when introduced into binary hydrogen-based structures of light elements, resulting in the formation of hydrogen alloy phases that exhibit high-temperature superconductivity, as demonstrated by the experimentally synthesized LaBeH$_8$ \cite{30}. Rare-earth elements often possess similar physical and chemical properties, giving rise to hydrogen-rich compounds with comparable cage-like hydrogen structures. However, subtle variations in their 4$f$ electron configurations can significantly influence their superconducting behaviors. Consequently, a comprehensive investigation of metal atom characteristics, including atomic size, valence electron configuration, electronegativity, and other relevant properties, is crucial for discovering new hydrogen-rich materials capable of achieving high-temperature superconductivity at moderate or even ambient pressures.

Recent advancements in research have continuously unveiled ternary hydrogen cage structures denoted as MXH$_1$$_2$. Intensive high-throughput computational screening and analysis have led to the identification of a range of high-temperature, stable, hydrogen-centric structures \cite{23, 31, 32, 33, 34, 35}. Our investigation expanded to analyze the ternary hydrogen-based structures with various guest metal atoms, focusing on both rare-earth and alkaline-earth elements. We identified nine compounds that exhibit stability within the pressure range of 50 $\sim$ 200 GPa. We thoroughly examined the metallic properties, dynamical stability, and electron-phonon ($e$-$ph$) coupling of these nine structures under varying pressures. Our computational findings indicate that ScYbH$_1$$_2$ and LuYbH$_1$$_2$ maintain dynamical stability above 70 GPa and ScLuH$_1$$_2$ above 100 GPa. Both CaLuH$_1$$_2$ and YLuH$_1$$_2$ are stable above 160 GPa. Significantly, CaLuH$_1$$_2$ exhibits a remarkably high superconducting critical temperature of 294 K at 180 GPa based on Eliashberg function calculations. This comprehensive analysis lays the groundwork for the discovery of new high-$T_{c}$ ternary hydride superconductors.


\section{Results and discussion }

\begin{figure}[t]
    \begin{center}
\includegraphics[width=18cm]{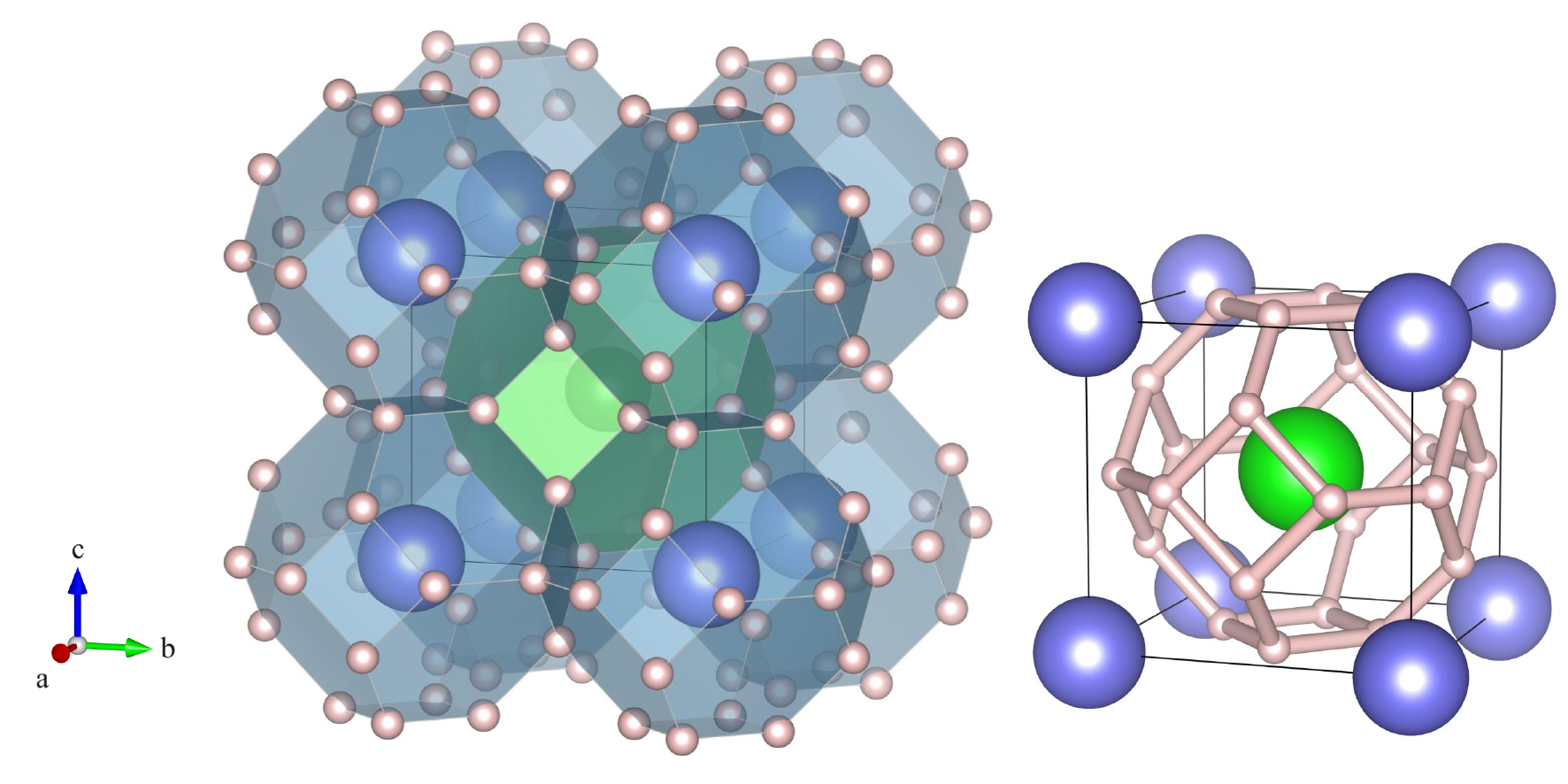}
\caption{\label{fig1} $Pm\overline{3}m$ structure of MXH$_{12}$, illustrated using the VESTA software\cite{48}. Purple, green, and pink spheres represent M, X, and H atoms, respectively.}
	\end{center}
\end{figure}

As the crystal structure of MXH$_1$$_2$ illustrated in Figure \ref{fig1}, the M, X, H atoms occupy the 1$a$ (0, 0, 0), 1$b$ (0.5, 0.5, 0.5) and 12$h$ (0.5, $y$, 0) Wyckoff positions. The H atoms form an interconnected framework, showing a sodalite-like cage, while the guest atom B occupies the central position of the hydrogen cage. Our extensive exploration of various combinations involving alkaline-earth elements from the  alkaline-earth and rare-earth elements has revealed nine stable hydrogen-based structures in the pressure range of 50 to 200 GPa. These structures include CaYH$_1$$_2$, CaScH$_1$$_2$, CaLuH$_1$$_2$, ScYH$_1$$_2$, CaYbH$_1$$_2$, YYbH$_1$$_2$, ScLuH$_1$$_2$, YLuH$_1$$_2$, LuYbH$_1$$_2$. Despite the varying metal atoms A and B, the nine computed inclusion compounds exhibit remarkable structural similarities, sharing the same space groups and exhibit similar lattice constants. This observation suggests that suitable atomic substitutions can preserve the structural stability of these compounds.

\begin{figure}[t]
\begin{center}
\includegraphics[width=18cm]{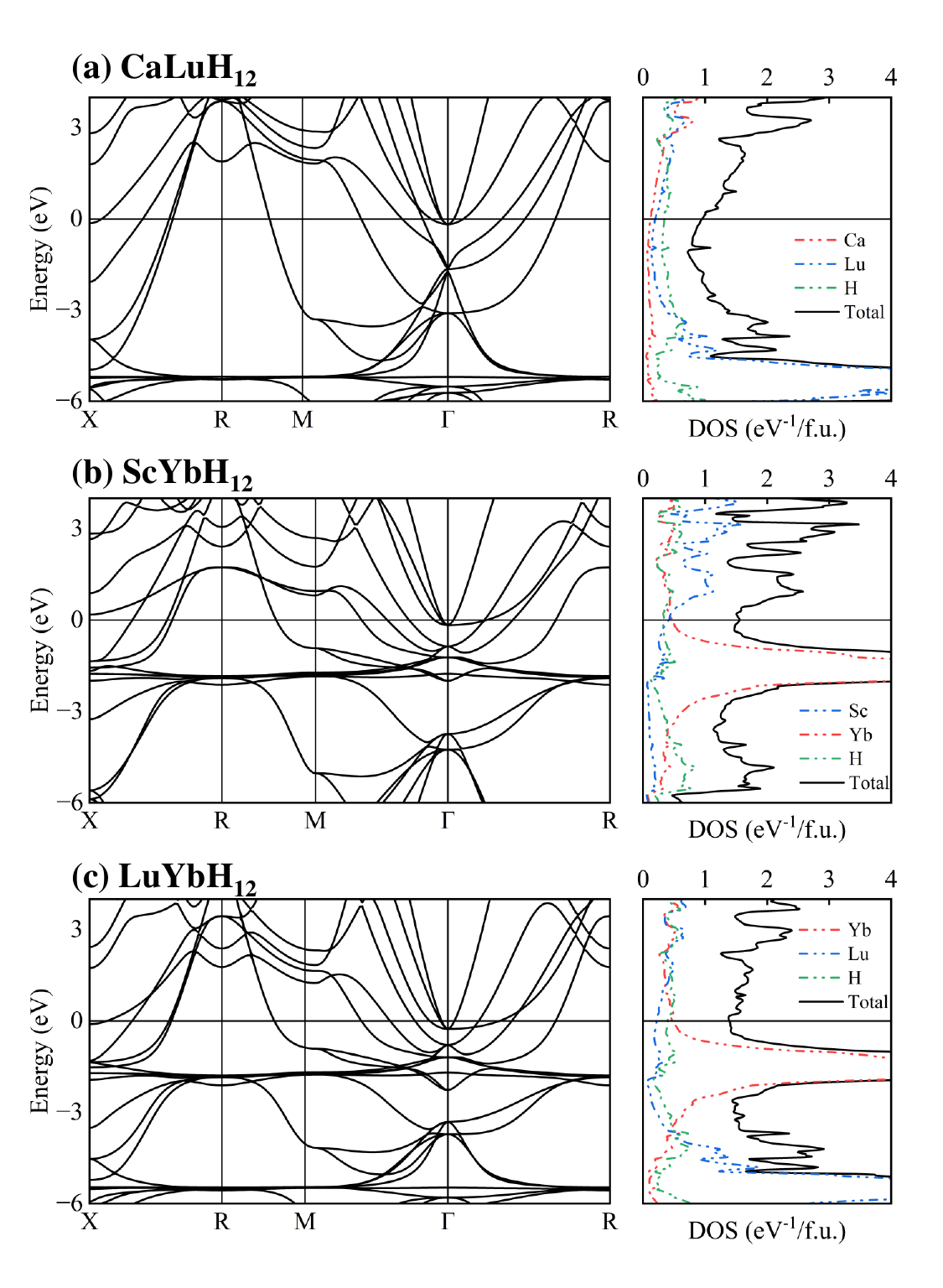}
\caption{ The electronic band structure and projected density of states of (a) CaLuH$_{12}$, (b) ScYbH$_{12}$, (c) YbLuH$_{12}$ at 160 GPa. The Fermi level is set as the reference energy level, with an energy value of zero.}
\label{fig2}
\end{center}
    \end{figure}

We calculated the electronic band structure and atomic projected density of states of $Pm\overline{3}m$-MXH$_1$$_2$ to elucidate the changes in electronic properties. We selected the electronic band structure of CaLuH$_{12}$, ScYbH$_{12}$ and LuYbH$_{12}$ at 160 GPa for comparison, as shown in Figure \ref{fig2}, and others are shown in the Supporting Information (SI). The band structure reveals numerous band crossings, resulting in intricate structural features predominantly governed by the crystal structure and associated symmetry operations. CaLuH$_{12}$, YLuH$_{12}$, and ScLuH$_{12}$ (Figure S1 in the SI) share the similar electronic structures and exhibit van Hove singularity (vHs) near the Fermi level ($E_f$). As the scandium element possesses one additional valence electron compared to calcium, and both scandium and yttrium belong to the same group  with similar electronic configurations, the compounds ScLuH$_{12}$ and YLuH$_{12}$ exhibit higher Fermi levels than CaLuH$_{12}$. For CaLuH$_{12}$, the bands observed in the energy range of -6 to -3.5 eV below the Fermi level, which are attributed to the electronic contributions from Lu atoms, correspond to the lower-energy states predominantly occupied by the $f$ orbitals of Lu. The density of states (DOS) distribution of H exhibits a relatively uniform character, suggesting a balanced participation of hydrogen atoms in the electronic structure without significant electron localization features. The outermost electron configuration of the Yb atom is 4$f^{14}$6$s^2$. The fully filled $f$ shell contributes significantly to enhancing compound stability. Consequently, the valence electron configuration of divalent ytterbium is 4$f^{14}$ \cite{36}. This electronic configuration is particularly evident in Figure \ref{fig2} (b), which illustrates the electronic structure of ScYbH$_{12}$. In this structure, the 4$f$ electrons of Yb predominantly occupy the energy band ranging from -2 to -1 eV, demonstrating the localized nature of these electrons and their contribution to the compound's stability. For the structures containing lutetium elements, the outermost electrons of Lu is 4$f^{14}$5$d^1$6$s^2$, and the extra electrons in the 5$d$ orbit cause the 4$f$ electrons to be far away from the Fermi level in the energy band structure. For LuYbH$_{12}$, the low-energy state is predominantly occupied by electrons provided by Lu, while the energy bands near the Fermi level is filled with electrons from Yb atoms. 

\begin{figure}
\begin{center}
\includegraphics[width=18cm]{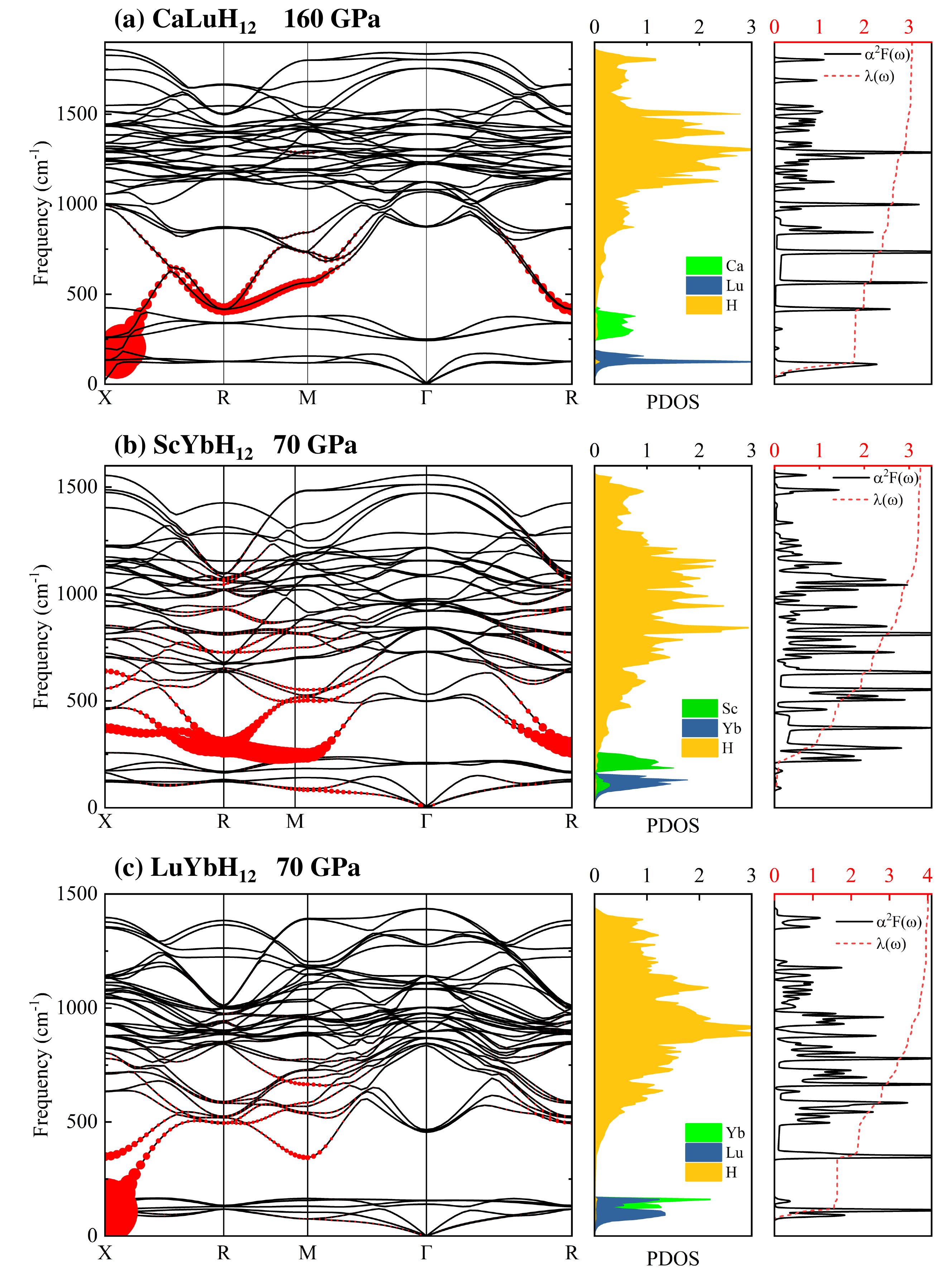}
\caption{Left panel: Phonon dispersion decorated with  the strength of $e$-$ph$ coupling indicated by red circles. Middle panel: phonon density of states, projections on M, X and H are shown in green, yellow and blue, respectively. Right panel: Eliashberg spectral function $\alpha^{2}F(\omega)$ in black solid line and the frequency-dependent $e$-$ph$ coupling constant $\lambda$($\omega$) with red dashed line.}
\label{fig3}
\end{center}
    \end{figure}

We calculated the phonon dispersion, phonon density of states (PDOS), Eliashberg spectral function $\alpha^{2}F(\omega)$  and $e$-$ph$ coupling parameter $\lambda$ of the studied MXH$_{12}$ system to explore the superconducting properties of different combinations. We selected the phonon spectra of CaLuH$_{12}$, ScYbH$_{12}$ and LuYbH$_{12}$ at the lowest pressure required to maintain structural stability, as shown in Figure \ref{fig3}. For CaLuH$_{12}$, The vibrational modes involving hydrogen atoms predominantly contribute to the phonon density of states (PDOS) over a wide frequency range of 500 $\sim$ 2000 cm$^{-1}$, as indicated by the yellow peaks in the PDOS plot. In the lower frequency range of  0 $\sim$ 500 cm$^{-1}$, PDOS is primarily dominated by vibrational modes involving the metal atoms A and B, represented by the blue and green peaks, respectively. The low-frequency region of the phonon spectrum, spanning from 0 to 250 cm$^{-1}$, arises primarily from the vibrational modes associated with the Lu or Yb atoms.

The maximum vibration frequency associated with hydrogen atoms is inversely correlated with the H-H bond length. Shorter H-H bonds result in higher corresponding vibration frequencies \cite{23}. Among our studied compounds, the closest H-H bond distance in the CaLuH$_{12}$ structure is 1.185 {\AA}, and the closest H-H bond distances in YbScH$_{12}$ and LuYbH$_{12}$ are 1.197 {\AA} and 1.247 {\AA} respectively. The phonon spectra of CaLuH$_{12}$, YbScH$_{12}$, and LuYbH$_{12}$ presented in Figure \ref{fig3} corroborate this relationship. The observed inverse correlation between bond length and vibration frequency in our results aligns with the proposed conclusion, providing validation for the accuracy of our calculations. The electron-phonon coupling strength is primarily concentrated in a few optical phonon branches, specifically the branches lie near 300 cm$^{-1}$ in the phonon dispersion. Electrons have an increased likelihood of interacting with specific phonon modes, which can modulate their scattering behavior and facilitate energy exchange between the electronic and vibrational systems. For LuYbH$_{12}$, the vibration modes of Lu and Yb are mainly concentrated in the range of 0 to 200 cm$^{-1}$. Between 200 cm$^{-1}$ and 400 cm$^{-1}$ in Figure \ref{fig3}(c), there is a phonon gap, resulting in a zero value of the Eliashberg function within this particular frequency range. At a frequency of 300 cm$^{-1}$, the joint action of the three elements in the structure leads to a prominent peak in the Eliashberg function. The $d$ or $f$ electrons in heavy metal elements usually form high-density electronic states in compounds, characterized by relatively large state electron density. This high-density electronic state enhances the interaction with phonons and affects the strength of $e$-$ph$ coupling. Above 500 cm$^{-1}$, the contribution of hydrogen to the parameter $\lambda$ dominates the value of the Eliashberg function. Hydrogen atoms, due to their small masses and atomic radius, can form strong interactions with other atoms. These interactions contribute significantly to the Eliashberg function, particularly in the high-frequency range above 500 cm$^{-1}$. However, it's important to note that the specific features of the Eliashberg function, such as peaks and contributions from different elements, are highly dependent on the electronic and lattice properties of the specific compound under consideration. In the low-frequency range between \(0\) and \(100\,\mathrm{cm}^{-1}\), the Eliashberg function exhibits negligible values, indicating a weak interaction between electrons and phonons. Consequently, this frequency region does not contribute significantly to the formation of Cooper pairs, a prerequisite for superconductivity. In contrast, for certain hydride compounds such as CaYbH$_{12}$ (Figure S2 (c) in SI) and ScYbH$_{12}$ (Figure \ref{fig3}(b)), the \(\lambda\) value, which is proportional to the strength of the electron-phonon coupling as described by the Eliashberg function, can attain significantly high values. As shown in Table~\ref{table1}, the \(\lambda\) value for CaYbH$_{12}$ reaches \(5.18\) at \(120\,\mathrm{GPa}\), while that for ScYbH$_{12}$ is \(3.25\) at \(70\,\mathrm{GPa}\).

\begin{figure}
\begin{center}
\includegraphics[width=18cm]{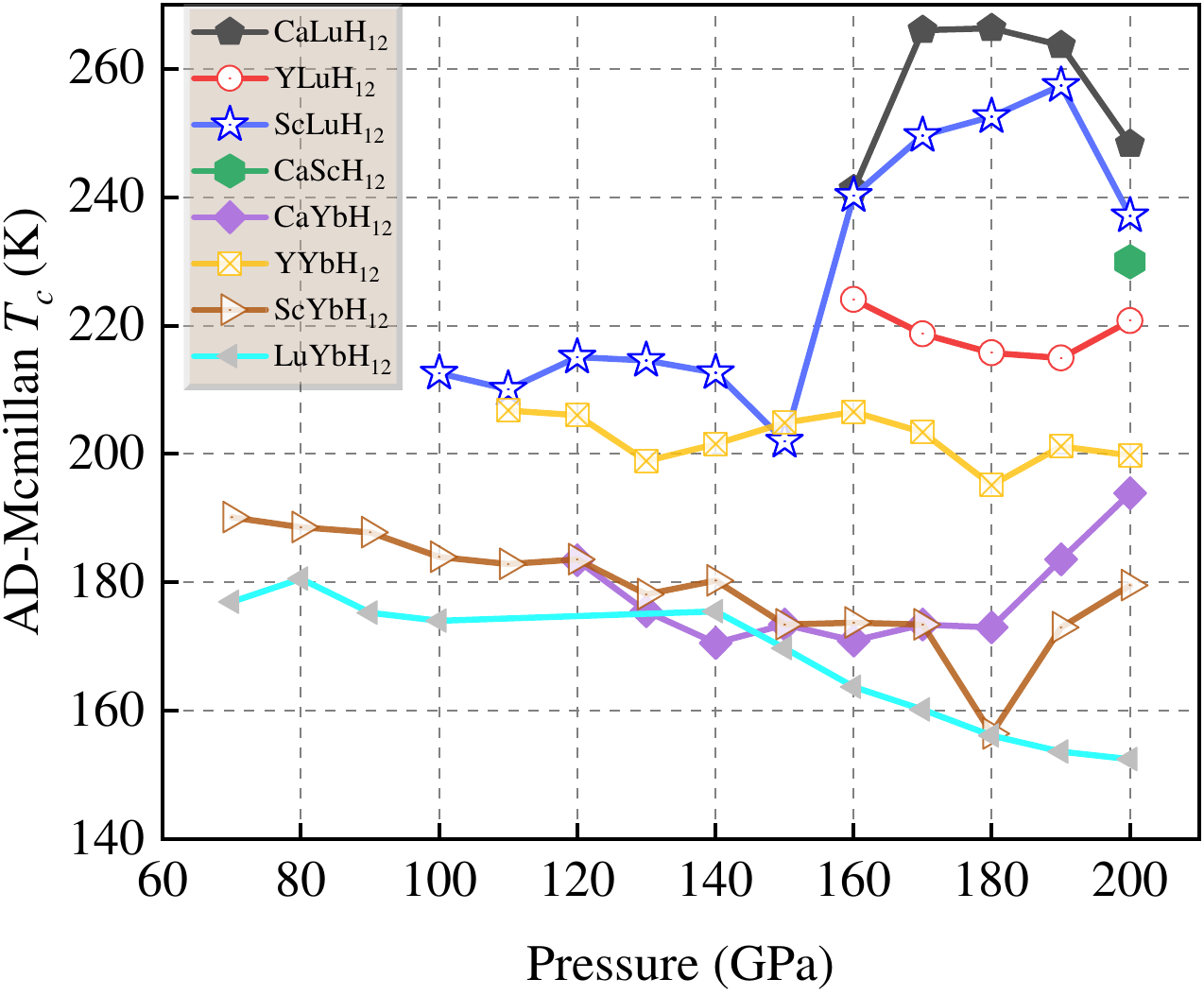}
\caption{The pressure dependence of $T_{c}$ calculated in the AD-McMillan equation of the $Pm\overline{3}m$ MXH$_{12}$ structures.}
\label{fig4}
\end{center}
    \end{figure}

Using a Coulomb pseudopotential parameter $\mu^*$= 0.10, we solved for the superconducting critical temperatures using the AD-McMillan and Eliashberg equations at different pressures, the results of which are presented in Table \ref{table1} and  Figure \ref{fig4}. Among the investigated compounds, CaLuH$_{12}$ exhibits a remarkable superconducting transition temperature of 294.2 K at 180 GPa, surpassing the highest predicted $T_{c}$ values for both CaH$_6$ \cite{26} and LuH$_6$ \cite{37}. In contrast, ScYbH$_{12}$ and LuYbH$_{12}$ demonstrate kinetic stability down to 70 GPa, while still exhibiting significantly high T$_{c}$ values. Among the MXH$_{12}$ compounds, ScYbH$_{12}$ and LuYbH$_{12}$ exhibit the lowest dynamic stable pressure, while ScYbH$_{12}$ has more advantages in $T_{c}$. Generally, an increase in pressure tends to augment lattice vibration frequencies, which in turn weakens the electron-lattice interactions, and the strength of $e$-$ph$ coupling typically dwindles with rising pressure. However, in compounds like ScLuH$_{12}$ and ScYbH$_{12}$, we observe an anomalous rise and fall in the coupling strength value, potentially attributable to alterations in the electronic band structure, which can significantly perturb the superconducting critical temperature as pressure varies. Furthermore, the incorporation of the Yb element can lower the pressure required for dynamic stability more effectively than the addition of the Lu element, owing to Yb's unique electronic structure; however, this substitution may weaken the superconducting transition temperatures.

\section{Conclusion}

In summary, we employed first-principles calculations to investigate the electronic structures, phonon properties, and $e$-$ph$ interactions of the $Pm\overline{3}m$ ternary hydrides CaLuH$_{12}$, YLuH$_{12}$, ScLuH$_{12}$, ScYbH$_{12}$, YYbH$_{12}$, CaYbH$_{12}$, and LuYbH$_{12}$ under varying pressure conditions. Through high-throughput computational screening, we discovered that several compounds in the MXH$_{12}$ system can attain dynamic stability at pressures as low as 70 GPa, concurrently exhibiting high superconducting transition temperatures. Remarkably, our calculations predict that CaLuH$_{12}$ could exhibit a remarkable $T_c$ of up to 294 K at 180 GPa, surpassing the highest predicted $T_c$ values for the binary hydrides CaH$_6$ and LuH$6$. The potential for high-temperature superconductivity in the MXH$_{12}$ system presents an opportunity and platform for exploring novel superconducting materials. These findings are expected to stimulate further theoretical and experimental research in the field of high-$T_c$ superconductors, particularly within the realm of ternary hydrides.

\section{Computational Methods}

We utilized our in-house developed crystal search algorithm software CRYSTREE to conduct a high-throughput structure search on MXH$_1$$_2$ in the pressure range of 50 $\sim$ 200 GPa \cite{38, 39}. CRYSTREE employs an extremal random forest regression (ERF) approach to accelerate the optimization process and efficiently generate potential candidate structures. Each iteration is preconfigured to yield 60 structures. Among the subsequently generated structures, 60$\%$ are obtained using the ERF method, while the remaining 40$\%$ are randomly generated. The calculation terminates when the difference in enthalpy values between the last 10 iterations falls below a specific energy threshold or reaches the maximum number of iteration steps that has been set. The full-potential linearized enhanced plane wave method WIEN2k of the Perdew-Burke-Ernzerhof (PBE) functional was used to calculate the electronic structure, obtain the electronic band and density of states \cite{40, 41}. We optimized the structure using the $ab$ $initio$ calculation method of the Quantum Espresso (QE) software package and calculated the $e$-$ph$ coupling parameter $\lambda$ and the logarithmic mean of the phonon vibration frequency $\omega$$_l$$_o$$_g$ for the stable compounds \cite{42}. Density functional perturbation theory (DFPT) was used to perform phonon calculations on the $Pm\overline{3}m$ structure under different pressures, including phonon dispersion and density of states \cite{43}. The charge density cutoff is 600 Ry and the wave function cutoff is 60 Ry. Among them, the pseudopotential is selected from the Standard Solid-State Pseudopotentials (SSSP) library \cite{44}. SSSP has been verified by the full electronic state equation and multiple standard tests of plane wave convergence of phonon frequency, band structure, cohesive energy and pressure. In the pseudopotential aspects of test standardization are widely recognized. Choose a 5$\times$5$\times$5 $q$-point grid and a 20$\times$20$\times$20 $k$-point grid, and use the optimized tetrahedral method to calculate the Brillouin zone integral and electron-phonon coupling. This method is applicable to both insulators and metals, where the elemental and band energies are linearized with respect to the wavevector $k$. A dense 32$\times$32$\times$32 grid is used to evaluate precise $e$-$ph$ coupling. In order to estimate the superconducting critical temperature of MXH$_1$$_2$ under pressure and to find the maximum $T_{c}$, we calculated the linear response of the $e$-$ph$ coupling based on the calculated the Eliashberg function $\alpha$$^2$$F$($\omega$) \cite{45}. The superconducting critical temperature is estimated using the Allen-Dynes modified McMillan (AD-McMillan) equations \cite{46, 47}. The equations are as follows:
\begin{equation}
{\alpha ^2}{\rm{F(}}\omega {\rm{) = }}\frac{1}{{2\pi N(0)}}\sum\limits_{Qv}^{} {\frac{{\gamma _{Qv}}}{{\omega _{Qv}}}\delta (\omega  - \omega _{Qv})},
\end{equation}
and
\begin{equation}
{T_c} = f_1f_2\frac{{{\omega _{log}}}}{{1.2}}{\rm{exp}}\left[ { - \frac{{1.04(1 + \lambda )}}{{\lambda  - {\mu ^*}(1 + 0.62\lambda )}}} \right],
\end{equation}
where $N$(0) is density of states at Fermi level, the partial contribution of each phonon mode to the total $e$-$ph$ coupling is proportional, where $\gamma_{Q\nu}$ is the $e$-$ph$ line width, and $\omega_{Q\nu}$ is the phonon frequency at phonon branch $\nu$ and wavevector $Q$. The larger is $\gamma_{Q\nu}$, the greater is the contribution to the Eliashberg spectral functions $\alpha$$^2$$F$($\omega$) and the $e$-$ph$ coupling constant $\lambda(\omega)$. 
\begin{equation}
\lambda(\omega)=2\int_{0}^{\omega}d\Omega\alpha^2F(\Omega)/\Omega
\end{equation}
The $\mu$$^*$ is the coulomb pseudopotential parameter, set to 0.1 in the calculation here. The logarithmic mean phonon frequency $\omega$$_l$$_o$$_g$ is defined as
\begin{equation}
{\omega _{\log }} = \exp [\frac{2}{\lambda }\int_0^\infty  {\frac{{d\omega }}{\omega }{\alpha ^2}{\rm{F}}(\omega )\ln \omega } ].
\end{equation}
The factor $f_{1}$ represents strong coupling correction, and the factor $f_{2}$ represents spectral function correction. The factors $f_{1}$ and $f_{2}$ depend on $\lambda$, $\mu^*$, $\omega_{log}$, and mean square frequency $\overline{\omega^2}$.
\begin{equation}
f_{1} = \sqrt[3]{1 + (\frac{\lambda }{{2.46(1 + 3.8{\mu ^*})}}) ^ {\frac{3}{2}}}
\end{equation}
\begin{equation}
f_{2}= 1-\frac{\lambda^2(1-\overline{\omega ^2}/\omega_{log})}{\lambda^2+3.312(1+6.3\mu^*)^2}
\end{equation}
\medskip
\textbf{Supporting Information}\\
Supporting Information is available from the Wiley Online Library or from the author.

\medskip
\textbf{Acknowledgement}\\
{This work is supported by the National Natural Science Foundation of China (Grants No. 12374135, 12175107), and NUPTSF (Grants No. NY219087, NY220038). Some of the calculations were performed on the supercomputer in the Big Data Computing Center (BDCC) of Southeast University.\\}

\textbf{Conflict of Interest}\\
The authors declare no conflict of interest.

\textbf{Data Availability Statement}\\
The data that support the findings of this study are available from the corresponding author upon reasonable request.

%

\newpage

\begin{table*}[!htbp]
    \centering
    \begin{tabular}{cccccccc}
    \hline\hline
          & Pressure (GPa) & $N$($E_f$) (states/Ry/f.u.) & $\lambda$ & $\omega _{\log}$ (K) & $\overline{\omega^2}$ (K) & AD-McMillan $T_{c}$ (K) & Eliashberg $T_{c}$ (K)\\
        \hline
        CaLuH$_{12}$ & 160 & 0.150 & 3.062 & 850 & 1297 & 241.2 & 238.7  \\ 
        ~ & 180 & 0.155 & 2.967 & 1063 & 1374 & 266.3 & 294.2  \\ 
        YLuH$_{12}$ & 160 & 0.207 & 2.930 & 822 & 1251 & 224.1 & 246.9  \\ 
        ScLuH$_{12}$  & 100 & 0.245 & 3.260 & 756 & 1045 & 212.6 & 231.4  \\ 
        ~ & 190 & 0.267 & 2.131 & 1414 & 1661 & 257.4 & 287.5  \\ 
        CaScH$_{12}$ & 200 & 0.212 & 3.311 & 754 & 1163 & 230.0 & 245.3  \\ 
        CaYbH$_{12}$  & 120 & 0.210 & 5.180 & 374 & 670 & 183.5 & 176.5 \\
        ~ & 200 & 0.215 & 1.780 & 1248 & 1512 & 193.9 & 218.0  \\ 
        YYbH$_{12}$ & 110 & 0.181 & 3.441 & 681 & 993 & 206.8 & 222.8  \\
        ScYbH$_{12}$  & 70 & 0.209 & 3.252 & 689 & 927 & 190.2 & 205.4  \\ 
        LuYbH$_{12}$ & 70 & 0.188 & 4.004 & 434 & 799 & 177.0 & 190.7 \\ 
        ~ & 80 & 0.191 & 2.986 & 708 & 936 & 180.6 & 198.4  \\ 
        CaYH$_{12}$ & 150 & 0.155 & 3.256 & 930 & 1228 & 253.7 & 281.1 \\ 
        \hline\hline
    \end{tabular}
    \caption{\label{table1} Electronic density of states at the Fermi level $N$($E_f$), $e$-$ph$ coupling constant $\lambda$, logarithmically averaged phonon frequency $\omega _{\log}$, mean square frequency $\overline{\omega^2}$, and critical temperature $T_{c}$ values estimated using Coulomb pseudopotential values of $\mu^*$=0.1 for different structures at different pressure.}
\end{table*}

\end{document}